\documentclass[aps,preprint,fleqn,prb,byrevtex,citeautoscript]{revtex4}

\usepackage{psfrag,graphicx,longtable}

\begin{document}  

\title{
Dependence of spontaneous emission rates of emitters on the refractive
 index of the surrounding medial
}

\author{Chang-Kui Duan}
\affiliation{
Institute of Applied Physics and College of Electronic Engineering,
Chongqing University of Posts and Telecommunications, Chongqing 400065, China.\\
}
\affiliation{Department of Physics and Astronomy, University of
  Canterbury, Christchurch, New Zealand}
\author{Michael F. Reid}
\affiliation{
Department of Physics and Astronomy \\ and MacDiarmid Institute of
Advanced Materials and Nanotechnology, University of Canterbury, Christchurch,
  New Zealand}

\date{\today}

\begin{abstract}
For emitters in a medium, different macroscopic or microscopic theoretical models
 predict substantially different dependencies of the spontaneous emission
 lifetime on refractive index. Various measurements have been carried out on Eu$^{3+}$,
 Ce$^{3+}$ and Nd$^{3+}$ ions, and quantum dots in different surrounding medial.
The dependence of the spontaneous emission rates on refractive index has been
interpreted with different models. By closely examining some of the experimental
results, we notice that some interpretations are based on implicit assumptions
 which are hard to justify, and some measurements contain too big uncertainties
to discriminate among different models. In this work we reanalyse the avilable measured
results and give a consistent interpretation.

\end{abstract}
\date{March 31, 2005}

\maketitle

\section{Introduction}

It is well known that the rate of spontaneous emission of emitters can be modified
by changing the surrounding dielectric medium.\cite{Blo1965,Top2003} The theory
on this subject continues to attract considerable attention due to 
its fundamental importance and its relevance to various
applications in low-dimensional optical materials and photonic
crystals\cite{Luk2002,Kum2003,Ber2004}. Various
macroscopic (see Ref. \onlinecite{Top2003} for a recent review) and
microscopic \cite{Bor1999,Cre2000,Ber2004} theoretical models have
been developed to model the dependence of the
spontaneous emission rates (or lifetimes) on refractive index. 
However, different models predict substantially different 
dependences of radiative lifetime on refractive index. 
There are also some measurements intended to discriminize these models.

In this paper: firstly, we summarize main theoretical models
and the corresponding underlying assumptions (Sec. II);  then
we examine the existing experimental results that have been claimed to
support different models and give them a consistent interpretations (Sec. III).

\section{Main theoretical models}
Four main models for the dependence of spantaneous emission rates for emitters
on the refractive index of the surrounding medium have been proposed since Purcell\cite{Pur1946}
noticed that the spnatenous emission rates can be modified by changing the environment.
Since most early theoretical and experimental results focus on the modification of the
rate by placing the emitters inside resonant cavities\cite{Kle1981,Mar1987}, 
and spatially inhomogeneous dielectrics
including photonic band gap materials\cite{Yab1987,Joh1994}. In those approaches the spantaneous emission
rate is proportionally altered when the radiation density of states is altered.
By quantizing the macroscopic Maxwell's equations, it can be seen that the density of states
of radiation is proportional to the refractive index $n$. Therefore one expression for the
radiative rate has been predicted by Nienhuis and Alkemad\cite{Nie1976} as
\begin{equation}
\Gamma_{\rm R}(n) = \Gamma_0 n,
\end{equation}
where $\Gamma_{\rm R}(n)$ and $\Gamma_0$ are the radiative decay rate of electric dipole
emitters in the dielectric with refractive index $n$ and in vacuum, respectively.
The drawback of expression Eq. (1) is that the dielectric is taken to be homogeneous 
over the entire space,  and also occupies the place occupied by emitters, so that the 
electric field felt by the emitter is equal to the macroscopic electric field 
(or proportional to the macroscopic electric field, with the coefficient not 
depending on the surrounding medium).

In general, the field felt by emitters is different from the macroscopic electric field.
Taking this difference into consideration leads to the so-called local-field corrections, which apply
to a wide variety of physical quantities including the spontaneous emission rate of
embedded emitters. If the system is isotropic on microscopic levels, then the local field
felt by the emitter is proportional to the macroscopic field with a ratio denoted as $f$, then we have
the corrected spontaneous emission rate for electric-dipole emitters\cite{Top2003}
\begin{equation}
\Gamma_{\rm R}(n) = \Gamma_o n f^2.
\end{equation}

Macroscopic derivations of the local-field correction often imply the use of
cavity around the emitter. The specific choice of the cavity is subtle mater, 
greatly complicating the interpretation of these models. One famous model
introduced by Lorentz have been used in many textbooks\cite{Jac1975,Kit1976}. 
A hypothetical spherical cavity filled with medium of the same average polarizability
 density as the surrounding dielectric is introduced in this model, but the
 contributions from the dipoles inside the cavity 
to the local electric field cancel. The ratio $f$ is given as
\begin{equation}
f_{\rm virtual} = \frac{n^2+2}{3}.
\end{equation}
The modification of radiative rate based on this ratio is usually called virtual-cavity
 model or Lorentz model\cite{Top2003}. Using the Lorentz model 
assumes that the polarizibility of the media is not altered by introducing the emitters
into the media. The results for Ce$^{3+}$ in various hosts with different
refractive indices support this models\cite{Dua2005a}.

In the case of fluorescent molecule emitters embedded in solution, 
the fluorescent molecules expel the solvent from the volume where they are
located and hence creat real cavities of dielectric there.
The ratio of the local electric field inside the cavity to the macroscopic field
 outside the cavity was given by Glauber and Lewenstein\cite{Gla1991} as
\begin{equation}
f_{\rm real} = \frac{3n^2}{2n^2+1}
 \end{equation}
The modification of electric-dipole radiative rate based on this ratio is usually called
 real-cavity model\cite{Top2003} or Glauber-Lewenstein model. There are quite a few 
experimental results\cite{Rik1995,Sch1998} on molecular emitters that support this model.

All these models described above are based on a macroscopic description
 of the dielectric. One
of the key features of these models is that the dielectric host are
 assumed to be unaffected by the presence of the embedded emitter.
 More recently, Crenshaw and Bowden\cite{Cre2000} proposed a fully microscopic model
 for local-field effects using quantum-electrodynamical, many-body
 derivation of Langevin-Bloch operator equatons for 
two-level emitters embedded in a dielectric host. The radiative rate
 can be written as
\begin{equation}
\label{microscopic}
\Gamma_{\rm R}(n) = \Gamma_0 \frac{n^2+2}{3}
\end{equation}
We shall refer to this model as Crenshaw-Bowden model.

Another very important result on this subject is the work by Berman
 and Milonni in early year 2004\cite{Ber2004}. They calculated the corrections to
 the electric-dipole decay rate by using another approach to the fully microscopic many-body
 theory. Their results are perfectly consistent with the macroscopic
 theories to the first order of dielectric density. However, to this
 order, their theory does not distinguish between the virtual-cavity
 and real-cavity models for the local-field correction. Therefore,
 there are still mainly four different models for
the radiative rate of embedded emitters, with two models 
compatible with both macroscopic and 
fully microscopic theories.

\section{Existing experimental results and reinterpretations}

\subsection{Reinterpretation of the lifetime of Y$_2$O$_3$:Eu$^{3+}$ 
nanoparticles in medium}
Recently we analyzed the lifetimes of $5d\rightarrow 4f$ transitions
of Ce$^{3+}$ ions in hosts of different refractive indices \cite{Dua2005c} and found that
virtual cavity model applies. We notice in the literature that there 
had been  experimental results on Y$_2$O$_3$:Eu$^{3+}$ nanoparticles embedded
in medium of various refractive indices\cite{Mel1999} which can also be interpreted with
virtual cavity model. However, these two situations are very different.
For the case of Ce$^{3+}$ in different hosts, Ce$^{3+}$ ions replace
cations of small polarizability and do not change the medium itself, so
that the virtual cavity model applies. However, for the case of
Y$_2$O$_3$:Eu$^{3+}$ nanoparticles embedded in medium, the medium is
expelled from the space occupied by the nanoparticle, which is similar
to the case of flurescent moleculars embedded in solutions. Therefore
the real-cavity model should apply, with the refractive index in the
equation replaced with the effective relative refractive index $n_r$ of the medium
(with effective refractive index $n_{\rm eff}(x)$) to the nanoparticle 
(assuming the bulk refractive index $n_{\rm Y_2O_3}$), {\it i.e.}, the radiative lifetime
\begin{equation}
\tau_{\rm R} = \frac{1}{\Gamma_{\rm R}}=\tau_{\rm bulk}\frac{1}{n_r} (\frac{2n_r^2+1}{3n_r^2})^2
\label{newmodel}
\end{equation}
where 
\begin{eqnarray}
&& n_r = \frac{n_{\rm eff}(x)}{n_{\rm Y_2O_3}}\\
&& n_{\rm eff} = x\cdot n_{\rm Y_2O_3} + (1-x)\cdot n_{\rm med}.
\end{eqnarray}
Here $n_{\rm med}$ is refractive index of the medium without Y$_2$O$_3$ particles,
and $x$ is the `filling factor' showing what fraction of space is occupied by the 
Y$_2$O$_3$ nanoparticles surrounded by the media.

\begin{figure}[htp]
\includegraphics[width = 16cm]{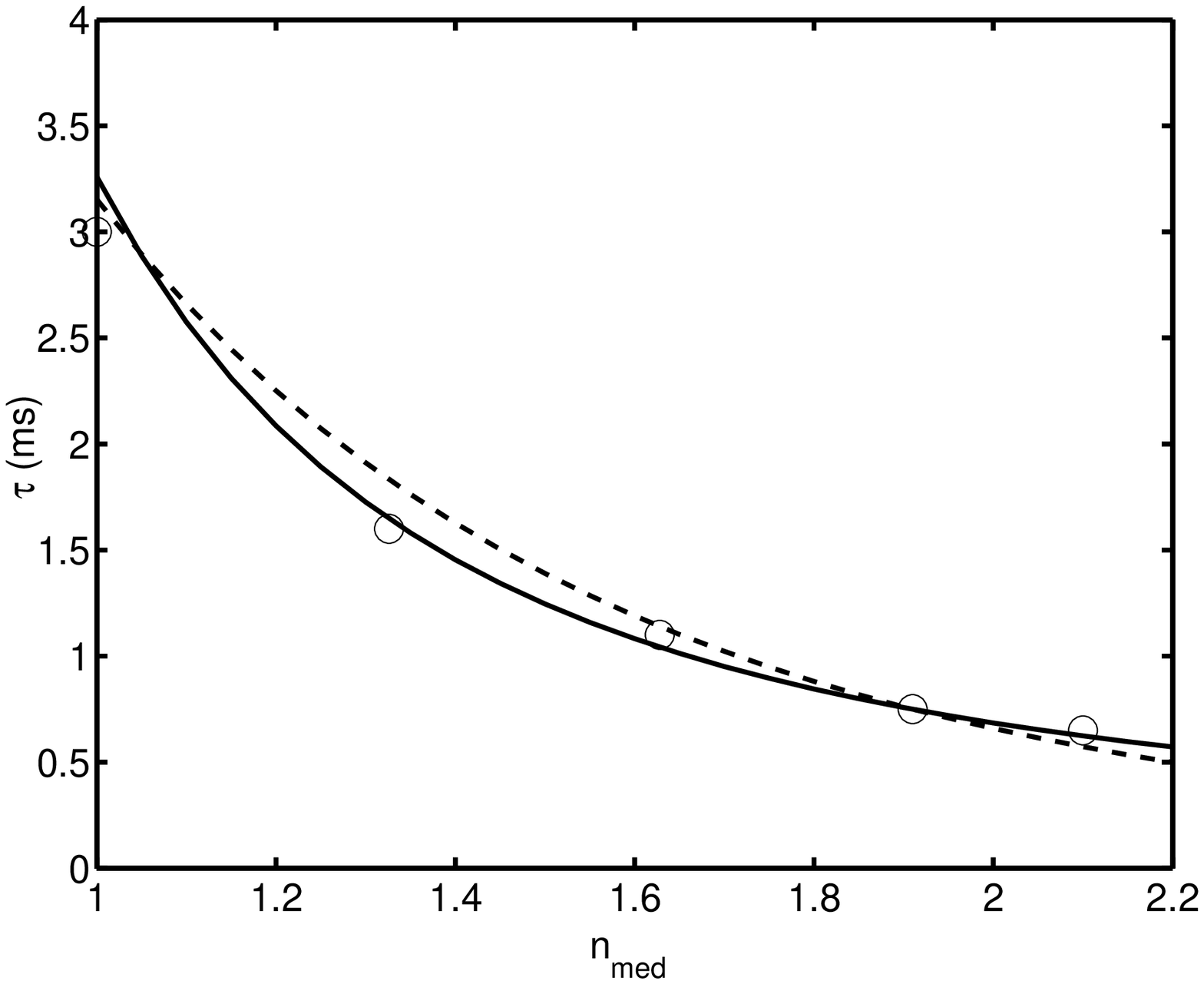}
\caption{
\label{fig1}
The dependence of the ${}^5D_0$ radiative lifetime $\tau_{\rm R}$ for the Eu$^{3+}$ 
C site on the refractive index of the media $n_{\rm med}$. 
Solid line: simulation with Eq. (\ref{newmodel}) using $x=0.15$; dashed line: simulated
with virtual-cavity model as given in FIG. 2 of Ref. \onlinecite{Mel1999}; circles:
experimental values as given in FIG. 2 of Ref. \onlinecite{Mel1999}.
}
\end{figure}

Fig. \ref{fig1} plots the radiative lifetime as a function of the medium.
It can be seen that Eq. (\ref{newmodel}) with filling factor $x=0.15$ fits
the measurements very well and hence provides an consistent alternative interpretation
to the original interpretation based on virtual-cavity model. The
filling factor is not explicitly measured and this value is also reasonable\cite{Mel1999}. 

\subsection{Reinterpretation of the lifetime of CdTe and CdSe quantum dots in media}
Very recently, the lifetime of CdTe and CdSe quantum dots has been measured and compared
with three of the above models\cite{Wui2004}. Although it is not conclusive, the fully microscopic 
model of Crenshaw and Bowden \cite{Cre2000} is suggested to best describe the measurements. There are
two possible problems with the interpretation: Firstly, as mentioned above, 
 in early year 2004 another group\cite{Ber2004} studied the corrections to radiative lifetime theoretically
using another approach of fully microscopic many-body theory and obtained
a theoretical result compatible with macroscopic models. Secondly, in the interpretation, the
measurement lifetime is considered as sololy due to radiative relaxation. It
is likely that part of the lifetime is due to intrinsic
 nonradiative relaxation. Actually, the measured quantum efficiency is around $55\%$ .

Here we give an alternative explanation of the correction to the lifetime.
Since the medium is  expelled from the space occupied by the quantum dots, 
the real-cavity model should apply. A reasonable assumption about the nonradiative
relaxation rate is that it does not depend on the medium. In such a case, the lifetime 
due to both radiative and nonradiative relaxation $\tau (n)$ 
and the quantum efficiency $\eta(n)$ of the quantum dots in medium can be written as
\begin{eqnarray}
&&\frac{1}{\tau(n)} = (\frac{3n^2}{2n^2+1})^2 n/\tau_{\rm R0}+ 1/\tau_{\rm NR},
\\
&&\eta(n) = 1.0-\frac{\tau(n)}{\tau_{\rm NR}}.
\end{eqnarray}

\begin{figure}[htp]
\includegraphics[width=16cm]{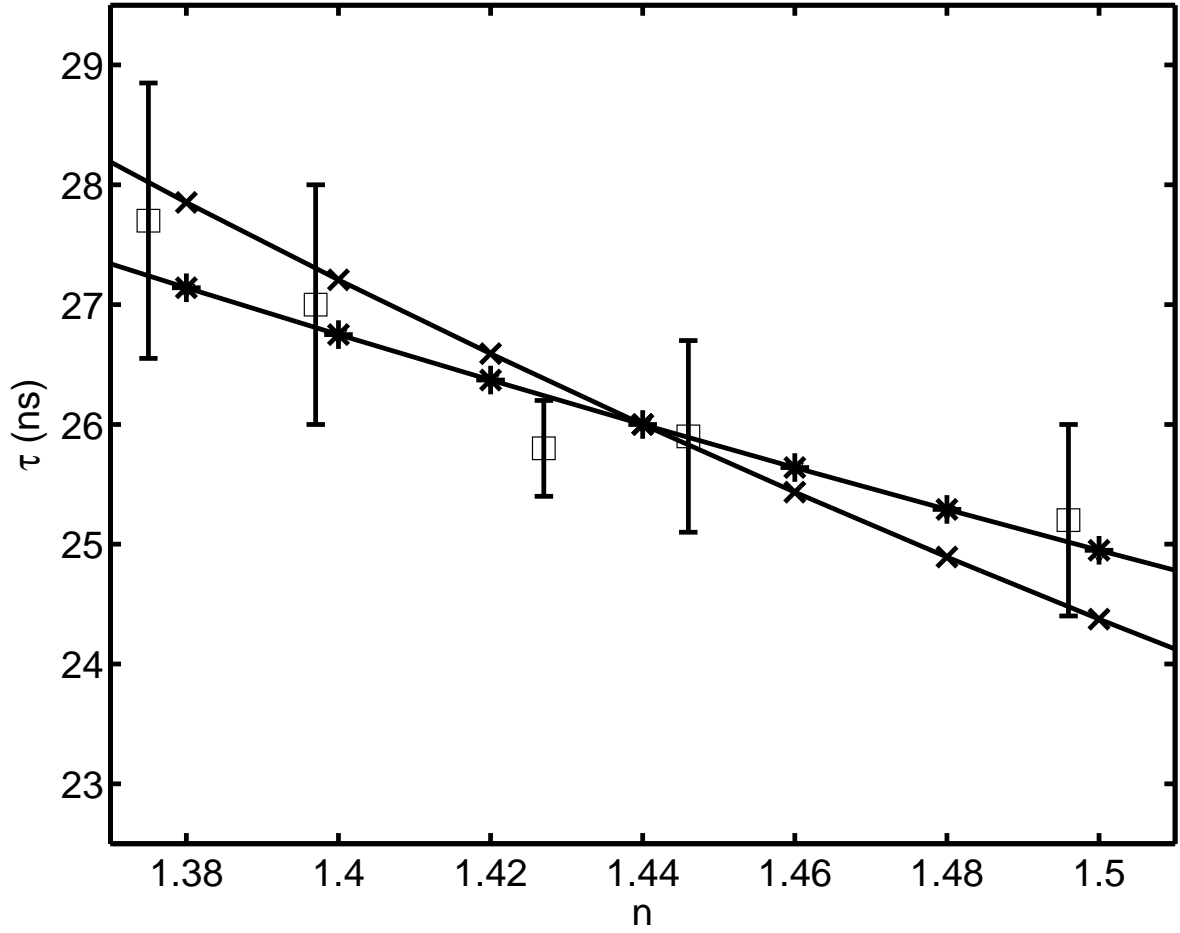}
\caption{
\label{fig2}
The dependence of the lifetime of excitons of GdTe quantum dots 
on the refractive index of solvent.
Squares: experimental data taken from FIG. 4 of Ref. \onlinecite{Wui2004};
triangles: simulated with Eq. (\ref{microscopic}) as given in 
 FIG. 4 of Ref. \onlinecite{Wui2004}; stars (overlaped with triangles)
and crosses: simulated with real-cavity model using best-fitted $\tau_{\rm NR} = 60 {\rm ns}$
(giving a quantum efficiency $\sim 57\%$ for $n\sim 1.4$-$1.5$) and a much higher
$\tau_{\rm NR} = 260 {\rm ns}$ (giving a quantum efficiency $\sim 90\%$.), respectively.
}
\end{figure}

Fig. \ref{fig2} plots the total lifetime of the quantum dot CdSe as a funciton of the
refractive index of the medium. Again, the real-cavity model
fits the experimental lifetime very well and give an quantum efficiency $57\%$ that
is is compatible with experiments.

\section{Summary}

In summary, experimental results for emitters in the form of nanoparticles 
or moleculars that expell the media and creat a real cavity, including those that
 are originally interpreted with other models, are compatible with the real-cavity
models. This is different from the case that emitters do not change the media,
 such as Ce$^{3+}$ ions replace cations with low polarizability,  where
the virtual-cavity model applies.

\section*{Acknowledgment}

C.K.D. acknowledges support of this work by the National Natural Science
Foundation of China, Grant No. 10404040 and 10474092.

\bibliography{lifetime,celifetime}

\clearpage

\end{document}